\documentclass[%
 reprint,
superscriptaddress,
 amsmath,amssymb,
 aps,
]{revtex4-2}

\usepackage{graphicx}% Include figure files
\usepackage{dcolumn}% Align table columns on decimal point
\usepackage{bm}% bold math
\usepackage[version=3]{mhchem}
\usepackage[nolist]{acronym}
\usepackage{siunitx}
\usepackage{fixltx2e}

\begin{document}

\preprint{APS/123-QED}

\title{Four-electron Negative-\textit{U} Vacancy Defects in Antimony Selenide}

\author{Xinwei Wang}
\affiliation{Thomas Young Centre and Department of Materials, Imperial College London, Exhibition Road, London SW7 2AZ, UK}
\author{Seán R. Kavanagh}
\affiliation{Thomas Young Centre and Department of Materials, Imperial College London, Exhibition Road, London SW7 2AZ, UK}
\affiliation{Thomas Young Centre and Department of Chemistry, University College London, 20 Gordon Street, London WC1H 0AJ, UK}
\author{David O. Scanlon}
\affiliation{Thomas Young Centre and Department of Chemistry, University College London, 20 Gordon Street, London WC1H 0AJ, UK}
\author{Aron Walsh}
\altaffiliation{a.walsh@imperial.ac.uk}
\affiliation{Thomas Young Centre and Department of Materials, Imperial College London, Exhibition Road, London SW7 2AZ, UK}
\affiliation{Department of Physics, Ewha Womans University, Seoul 03760, Korea}

\date{\today}% It is always \today, today,

\begin{abstract}
The phenomenon of negative-\textit{U} behavior, where a defect traps a second charge carrier more strongly than the first, has been established in many host crystals. Here we report the case of four-carrier transitions for both vacancy defects in \ce{Sb2Se3}. A global structure searching strategy is employed to explore the defect energy landscape from first-principles, revealing previously-unrealized configurations which facilitate a major charge redistribution. Thermodynamic analysis of the accessible charge states reveals a four-electron negative-\textit{U} transition ($\Delta q$ = 4) for both V$_\textrm{Se}$ and V$_\textrm{Sb}$ and, by consequence, amphoteric behavior for \emph{all} intrinsic defects in \ce{Sb2Se3}, with impact on its usage in solar cells. To the best of our knowledge, four-electron negative-\textit{U} behavior has not been previously explored in this or other compounds. The unusual behavior is facilitated by valence alternation, a reconfiguration of the local bonding environments, characteristic of both Se and Sb. 
% UPLOAD
% The phenomenon of negative-U behavior, where a defect traps a second charge carrier more strongly than the first, has been established in many host crystals. Here we report the case of four-carrier transitions for both vacancy defects in Sb2Se3. A global structure searching strategy is employed to explore the defect energy landscape from first-principles, revealing previously-unrealized configurations which facilitate a major charge redistribution. Thermodynamic analysis of the accessible charge states reveals a four-electron negative-U transition (delta q = 4) for both V_Se and V_Sb and, by consequence, amphoteric behavior for all intrinsic defects in Sb2Se3, with impact on its usage in solar cells. To the best of our knowledge, four-electron negative-U behavior has not been previously explored in this or other compounds. The unusual behavior is facilitated by valence alternation, a reconfiguration of the local bonding environments, characteristic of both Se and Sb. 
\end{abstract}

\maketitle

\begin{acronym}
\acro{PV}{photovoltaic}
\acro{PCE}{power conversion efficiency}
\acro{TAS}{thermal admittance spectroscopy}
\acro{DLTS}{deep-level transient spectroscopy}
\acro{ODLTS}{optical deep-level transient spectroscopy}
\acro{1D}{one-dimensional}
\acro{3D}{three-dimensional}
\acro{TL}{transition level}
\acro{CBM}{conduction band minimum}
\acro{vdW}{van der Waals}
\acro{VASP}{Vienna Ab initio Simulation Package}
\acro{DFT}{density functional theory}
\acro{PES}{potential energy surface}
\end{acronym}

%【INTRODUCTION】

Point defects are unavoidable and play a decisive role in the performance of semiconductor devices\cite{pantelides1978electronic}. They are localized in real space and can act as charge traps. A defect may be able to capture two charge carriers of the same type with a larger binding energy for the second than the first, if the stabilization energy (e.g. from structural relaxation and exchange interactions) compensates the on-site Coulomb repulsion energy between the two carriers\cite{boer2018semiconductor,coutinho2020characterisation}. This energy cost of trapping an \emph{additional} charge carrier is defined as the Hubbard correlation energy (\textit{U})\cite{hubbard1963electron}, and such a defect with negative correlation energy is termed a negative-\textit{U} center. The concept of negative-\textit{U} was first proposed by Anderson \cite{anderson1975model} to explain the preference of paired electrons in chalcogenide glasses and expanded by Street and Mott\cite{street1975states} to defects in the same materials. Kastner et al. \cite{kastner1976valence} then proposed a valence-alternation model to account for the atomic reconstructions arising from dangling bonds. They noted that valence alternation is applicable to both amorphous and crystalline chalcogenides as well as materials with group-V atoms (e.g. As, Sb, Bi). Since then, negative-\textit{U} behavior has been widely reported in many systems beyond chalcogenides (including oxides, carbides, silicon and compound semiconductors such as GaAs and CdTe).\cite{coutinho2020characterisation,baraff1979silicon,troxell1980interstitial,hemmingsson1998negative,alt1990experimental}.

Antimony selenide (\ce{Sb2Se3}) has emerged as an earth-abundant and environmental-friendly absorber layer for thin-film solar cells due to its promising electronic and optical properties\cite{wang2022lone}. Despite remarkable progress made in improving its \ac{PCE} over the last decade, the record efficiency is \SI{10.57}{\percent}\cite{zhao2022regulating}, which is still far from the detailed-balance limit of $\sim$\SI{30}{\percent} and performance of other commercial solar cells.
A key inhibitor of device performance in \ce{Sb2Se3} is defect-assisted recombination\cite{dong2021boosting}.
Point defects in \ce{Sb2Se3} have been widely studied by experimental measurements\cite{lian2022distinctive,hu2018investigation,wen2018vapor,hobson2020defect} and first-principles calculations \cite{liu2017enhanced,savory2019complex,huang2019complicated,stoliaroff2020deciphering,huang2021more}. Defect-detection techniques such as \ac{TAS}, \ac{DLTS} and \ac{ODLTS} are able to provide information on defect density and energy levels, whereas the identification of defect atomic identity relies heavily on the theoretically calculated thermodynamic \ac{TL}\cite{freysoldt2014first}. Thus, accurate prediction of defect behavior and therefore \ac{TL} position is essential. 

Within the standard procedure of simulating point defects in solids, defect structures are generated simply by removing, adding or substituting one atom in an otherwise pristine supercell, and keeping all other atoms fixed. Then starting from these undistorted initial configurations, gradient-based structural relaxation is performed to obtain the equilibrium defect geometries.
However, this approach will find the local minimum configuration \emph{closest} to the high-symmetry initial structure on the \ac{PES}, which may not be true ground-state defect structure\cite{mosquera2021search}. Metastable structures are thus obtained and the predicted properties are severely affected. The widespread prevalence of this issue has recently been demonstrated, and is expected to be greatly exacerbated for systems such as \ce{Sb2Se3} with low-symmetry crystal structures and flexible bonding environments, which yield complex \ac{PES}s with many local minima\cite{kavanagh2022impact}.

\begin{figure*}[ht]
    \centering
    {\includegraphics[width=0.75\textwidth]{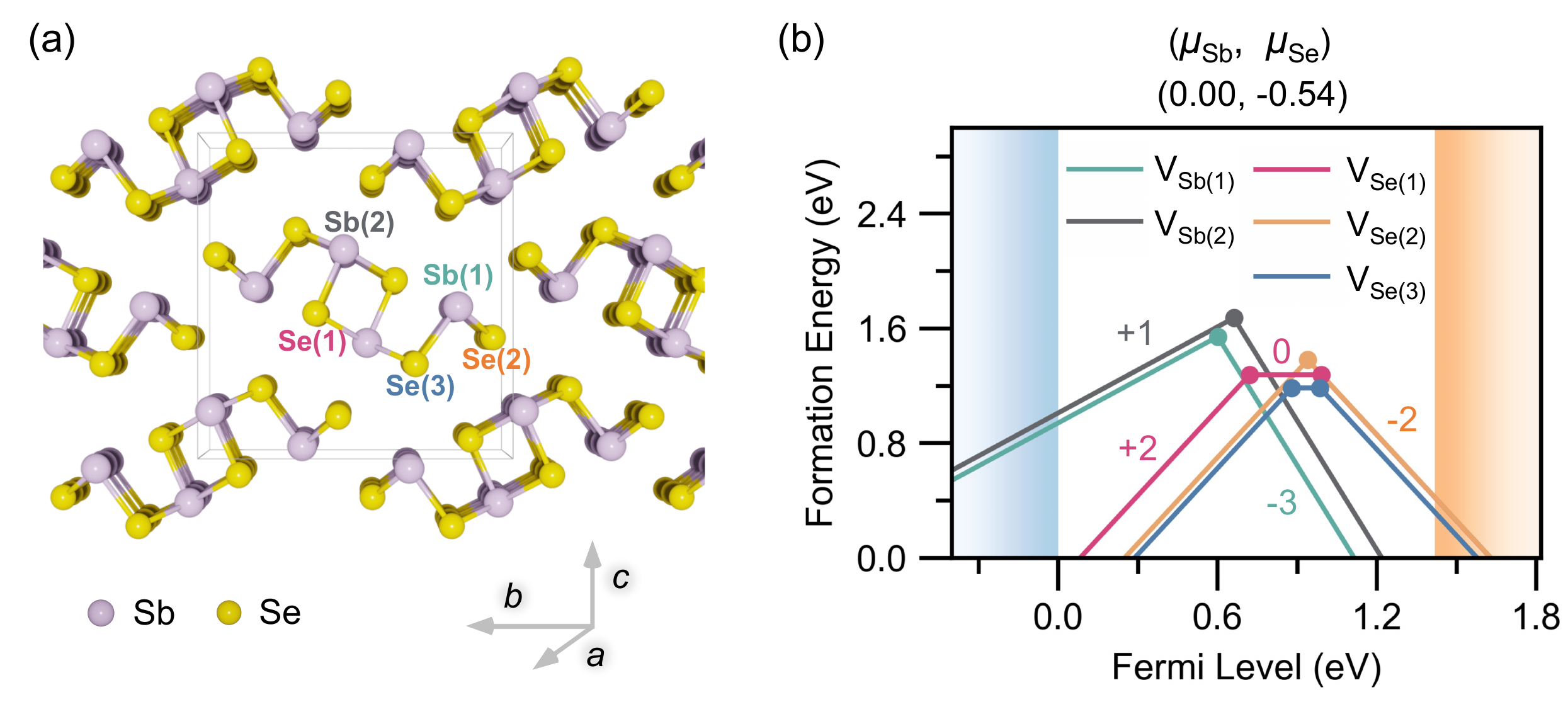}} \\
    \caption{(a) Ground-state crystal structure (\textit{Pnma} space group) of \ce{Sb2Se3}. The conventional unit cell is represented by a cuboid. (b) Defect formation energy diagram of V$_\textrm{Se}$ and V$_\textrm{Sb}$ under Sb-rich equilibrium growth condition. Charge states are shown adjacent to lines, and transition levels are given by filled circles. Valence band maximum (VBM) in blue, set to \SI{0}{eV}, and conduction band in orange.}
    \label{fig_structure}
\end{figure*}

In this Letter, motivated by the observations of valence alternation in chalcogenide semiconductors and the recent development of the \textsc{ShakeNBreak}\cite{mosquera2022shakenbreak} global \ac{PES} search method, we have reinvestigated the accessible charge states and structures of vacancies in \ce{Sb2Se3}. We found that previous theoretical studies \cite{liu2017enhanced,savory2019complex,huang2019complicated,stoliaroff2020deciphering,huang2021more} considered only a limited set of possible charge states (corresponding to conventional defect behaviors) and followed the standard defect modeling approach described above, without a comprehensive search for low-energy structural configurations. We show that this results in an incomplete understanding of defects in \ce{Sb2Se3}, failing to identify the valence alternation and strong self-compensating behaviour of vacancies.
Specifically, V$_\textrm{Sb}$ has been reported to be a shallow acceptor with accessible 0, -1, -2 and -3 charge states, and V$_\textrm{Se}$ a deep donor with 0, +1 and +2 charge states\cite{liu2017enhanced,savory2019complex,huang2019complicated,stoliaroff2020deciphering,huang2021more}. 
Here we reveal that both V$_\textrm{Sb}$ and V$_\textrm{Se}$ are in fact amphoteric, with unconventional stable positive charge states for V$_\textrm{Sb}$ and negative charge states for V$_\textrm{Se}$. Moreover, we predict four-electron negative-\textit{U} transitions for both Sb and Se vacancies.

%【METHODS】 
The equilibrium geometric and electronic structures of vacancy defects in \ce{Sb2Se3} were calculated within the framework of Kohn-Sham \ac{DFT}\cite{kohn1965self,dreizler1990density} as implemented in the \ac{VASP}\cite{kresse1996efficient}. 
The projector augmented-wave (PAW) method\cite{kresse1999ultrasoft} was employed with a converged plane-wave energy cutoff of 350 eV.
The Heyd-Scuseria-Ernzerhof hybrid functional (HSE06)\cite{heyd2003hybrid,krukau2006influence} and D3 dispersion correction\cite{grimme2004accurate} were used for both the structural relaxation and the static calculation of total energy.
The atomic positions were optimised until the Hellman-Feynman forces on each atom were below \SI{0.01}{eV/atom}.
Trial defect configurations were obtained using \textsc{ShakeNBreak}\cite{mosquera2022shakenbreak}, in 3$\times$1$\times$1 (\SI{11.86}{\angstrom}$\times$\SI{11.55}{\angstrom}$\times$\SI{11.93}{\angstrom}) supercells of the conventional unit cell. Here, chemically-guided local bond distortions around the defect were employed, alongside random perturbation to all atoms in the supercell, in order to sample the defect \ac{PES} and identify potential energy-lowering reconstructions. The workflow and details are provided in Section S1.
The defect formation energy of a defect \textit{D} in charge state \textit{q} is defined as\cite{zhang1991chemical,freysoldt2014first}:
\begin{equation}
    \Delta E^{\textit{f}}_{D,q} = E_{D,q} - E_\textrm{host} + \sum_{i}n_{i}\mu_{i} + qE_F +E_\textrm{corr}
\end{equation}
where $E_{D,q}$ and $E_\textrm{host}$ are the total energies of the supercell with and without the defect \textit{D}, respectively. $n_i$ and $\mu_{i}$ indicate the number and the chemical potential of added ($n_i$ \textless { }0) or removed ($n_i$ \textgreater { }0) atom of type \textit{i}, respectively. $E_F$ is the Fermi level and $E_\textrm{corr}$ is a correction term for spurious interactions between charged defects under periodic boundary conditions. The correction scheme developed by Kumagai and Oba\cite{kumagai2014electrostatics} which includes anisotropic dielectric screening is used in this work.

%【RESULTS】
\textit{Accessible vacancy charge states in \ce{Sb2Se3}}.
As shown in Fig. \ref{fig_structure}(a), the ground-state crystal structure of \ce{Sb2Se3} is orthorhombic with quasi-\ac{1D} [Sb$_4$X$_6$]$_n$ ribbons stacked together via weak interactions\cite{wang2022lone}. The unique crystal structure makes it more tolerant to large local lattice deformation compared to conventional crystals with \ac{3D} connectivity. Due to the low crystal symmetry, there are two inequivalent Sb sites and three inequivalent Se sites (shown in Fig. \ref{fig_structure}(a)), with all sites considered in this work.
The defect formation energy diagram in Fig. \ref{fig_structure}(b) plots the thermodynamically stable charge state as a function of Fermi level in the band gap for vacancy defects in \ce{Sb2Se3}. As expected, the multiple inequivalent sites with different local environments leads to small but significant differences in the properties of defects of the same type. In line with previous studies \cite{liu2017enhanced,savory2019complex,huang2019complicated,stoliaroff2020deciphering,huang2021more}, 0 and +2 charge states are found to be stable for V$_\textrm{Se}$, and -3 for V$_\textrm{Sb}$. On the other hand however, abnormal charge states of -2 for V$_\textrm{Se}$ and +1 for V$_\textrm{Sb}$ are also found, revealing strong amphoteric behaviour for both V$_\textrm{Se}$ and V$_\textrm{Sb}$.
In consequence, a (+1/-3) charge transition level is observed lying in the middle of the gap for V$_\textrm{Sb}$, regardless of Sb site, indicating that a change in Fermi level will lead to capture or release of four electrons per vacancy. Likewise, a four-electron (+2/-2) \ac{TL} is found for V$_\textrm{Se(2)}$, while V$_\textrm{Se(1)}$ and V$_\textrm{Se(3)}$ have (+2/0) and (0/-2) \ac{TL}s. Thus both V$_\textrm{Sb}$ and V$_\textrm{Se}$ are amphoteric negative-\textit{U} centers, with more typical two-electron negative-\textit{U} levels in V$_\textrm{Se(1)}$ and V$_\textrm{Se(3)}$ (though exhibiting two such levels each), but rare four-electron transfer levels for V$_\textrm{Sb(1)}$, V$_\textrm{Sb(2)}$ and V$_\textrm{Se(2)}$. Notably, amphoteric behavior has previously been reported for antisite and interstitial defects in \ce{Sb2Se3}\cite{savory2019complex,huang2021more}, but not for vacancies due to the requirement of structure-searching methods as mentioned above. Thus, our results show that \emph{all} intrinsic defects in \ce{Sb2Se3} are in fact amphoteric, having the ability to capture both electrons and holes, and yield strong ionic charge compensation. These properties have important consequences for defect-mediated carrier trapping, recombination and dopability in \ce{Sb2Se3}, which are key factors for photovoltaic device performance.

\textit{Electronic and structural reconfigurations}.
Negative-\textit{U} behavior at defects is often closely related to lattice distortion and structural reconstruction\cite{street1975states,coutinho2020characterisation,kavanagh2021rapid}. Before examining the defect configurations in detail, let us understand the electronic structure of \ce{Sb2Se3} first.
In pristine \ce{Sb2Se3}, the oxidation states of Sb (5$s^{2}5p^3$) and Se (4$s^{2}4p^4$) are +3 and -2, respectively, as each Sb cation donates three electrons to the neighboring Se anions, and each Se anion accepts two electrons from the neighboring Sb cations on average. 
Upon formation of a neutral vacancy defect by removing an atom and its valence electrons, dangling bonds with excess charge in the form of electrons/holes are introduced. Specifically, the removal of Sb with its three donated electrons to create V$^{0}_{\textrm{Sb}}$ results in three holes on three Se dangling bonds, while removing Se leaves behind its two accepted electrons with two Sb dangling bonds in V$^{0}_{\textrm{Se}}$.     
In the fully-ionized state (i.e. V$^{3-}_{\textrm{Sb}}$ and V$^{2+}_{\textrm{Se}}$), electrons are added/removed from the defect site to remove the excess charge and compensate the dangling bonds, stabilizing this charge state. Other charge states can also be stabilized, by localizing the excess charge (e.g. bound small polarons) and/or through atomic rearrangements to form compensating bonds and accommodate the excess charge\cite{kavanagh2021rapid}. 

\begin{figure}[ht]
    \centering
    {\includegraphics[width=0.5\textwidth]{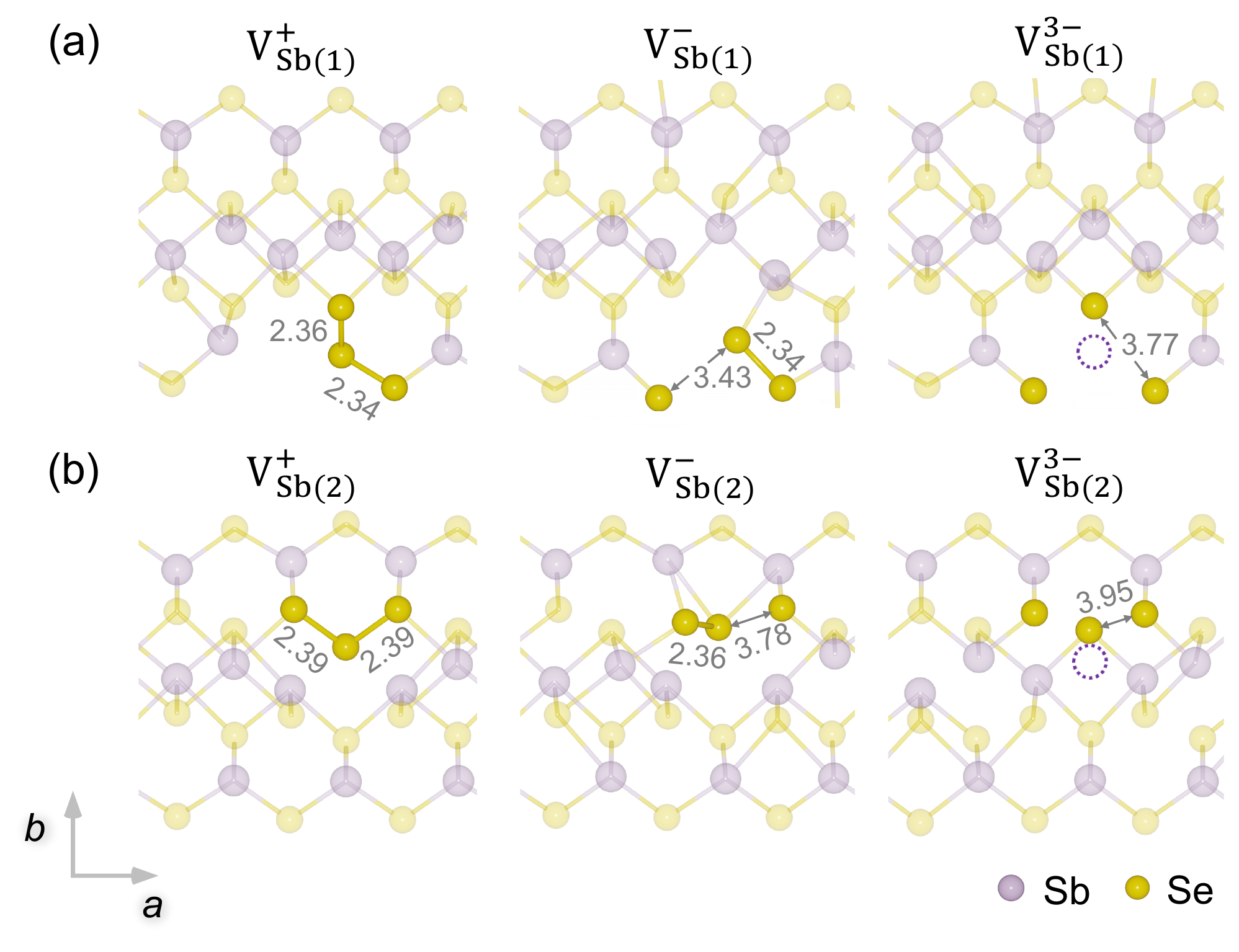}} \\
    \caption{The local defect geometry of V$_\textrm{Sb}$ as a function of charge state. The subscripts (1) and (2) refer to two inequivalent sites. The Se-Se bond lengths are shown in Å, and the vacant Sb site is denoted by a dotted circle.}
    \label{fig_vsb}
\end{figure}

\begin{figure}[ht]
    \centering
    {\includegraphics[width=0.5\textwidth]{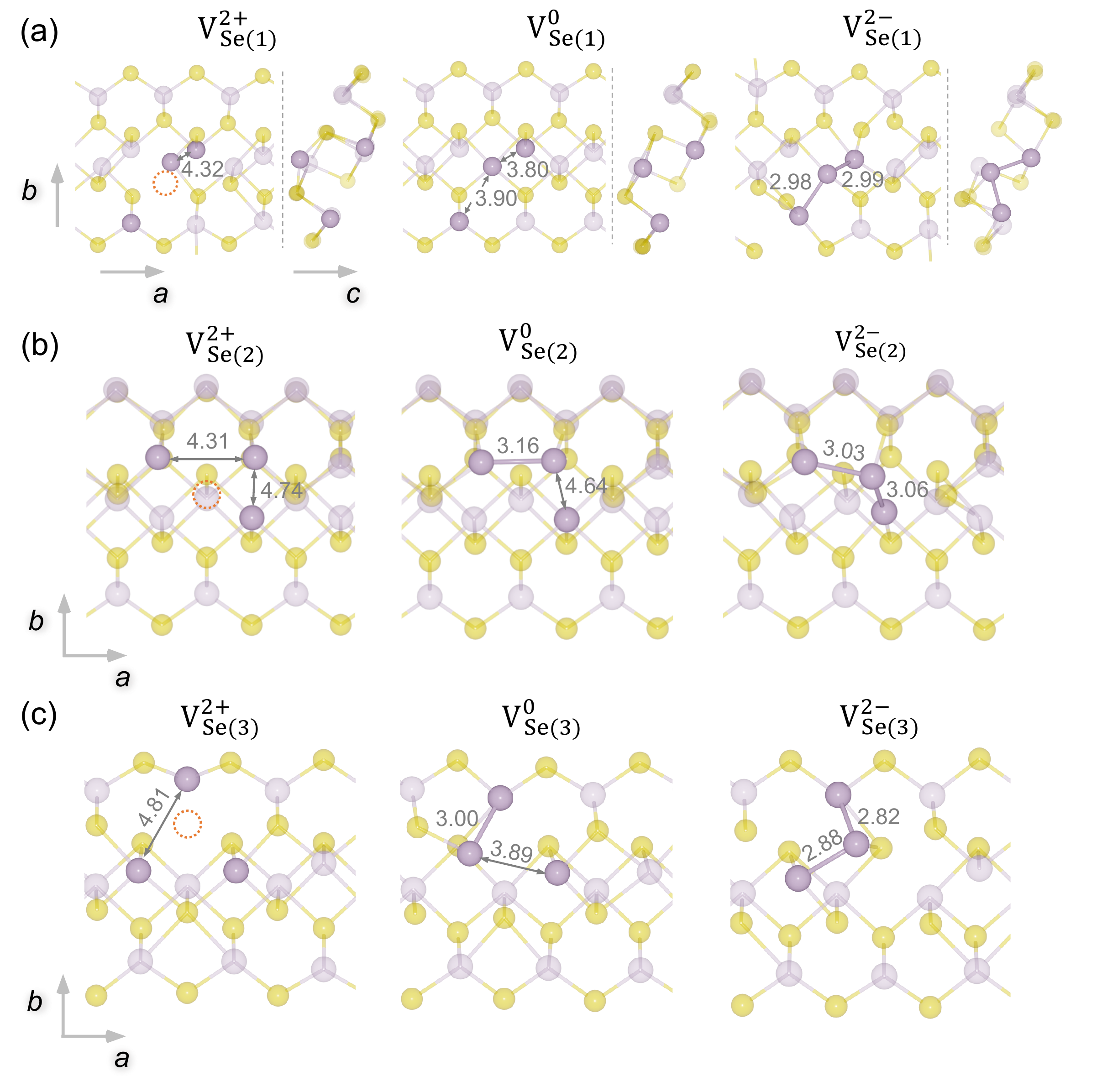}} \\
    \caption{The local defect geometry of V$_\textrm{Se}$ as a function of charge state. The subscripts (1), (2) and (3) refer to inequivalent sites. The Sb-Sb bond lengths are shown in Å, and the vacant Se site is denoted by a dotted circle.}
    \label{fig_vse}
\end{figure}

Taking V$_\textrm{Sb}$ as an example, due to the flexibility of the crystal structure and the tendency to electron pairing for Se species\cite{kastner1976valence}, alternative V$_\textrm{Sb}$ charge states can be stabilized by the migration of Se anions to form compensating Se-Se bonds at the vacancy. By eliminating the excess holes with this sharing of electrons between dangling bonds, the system energy is lowered. 
As shown in Fig. \ref{fig_vsb}, the two excess holes in V$^{-}_{\textrm{Sb}}$ are filled by the formation of a Se-Se dimer, while a Se-Se-Se trimer is formed in V$^{+}_{\textrm{Sb}}$ to pair all electrons and fill the four excess holes for this species.
V$^{+}_{\textrm{Sb}}$ can thus be thought of as a V$^{2+}_\textrm{Se}$-Se$^{-}_\textrm{Sb}$ complex.
Such structural reconstruction and negative-\textit{U} behavior due to paired dangling bonds has previously been termed valence alternation\cite{kastner1976valence}, with similar re-bonding behavior reported in lone-pair chalcogenide systems\cite{street1975states,kastner1976valence,kolobov1996origin,boer2018semiconductor}.
Similarly, a Sb-Sb-Sb trimer is also found in V$^{2-}_{\textrm{Se}}$, regardless of Se site (Fig. \ref{fig_vse}). 
The formation of an antisite originated from a vacancy in \ce{Sb2Se3} is also supported by experimental evidence\cite{che2022thermal}, which has found that post-annealing treatment facilitates the transformation from V$_\textrm{Se}$ to Sb$_\textrm{Se}$ by the migration of neighbouring Sb cations.

\begin{figure}[ht]
    \centering
    {\includegraphics[width=0.48\textwidth]{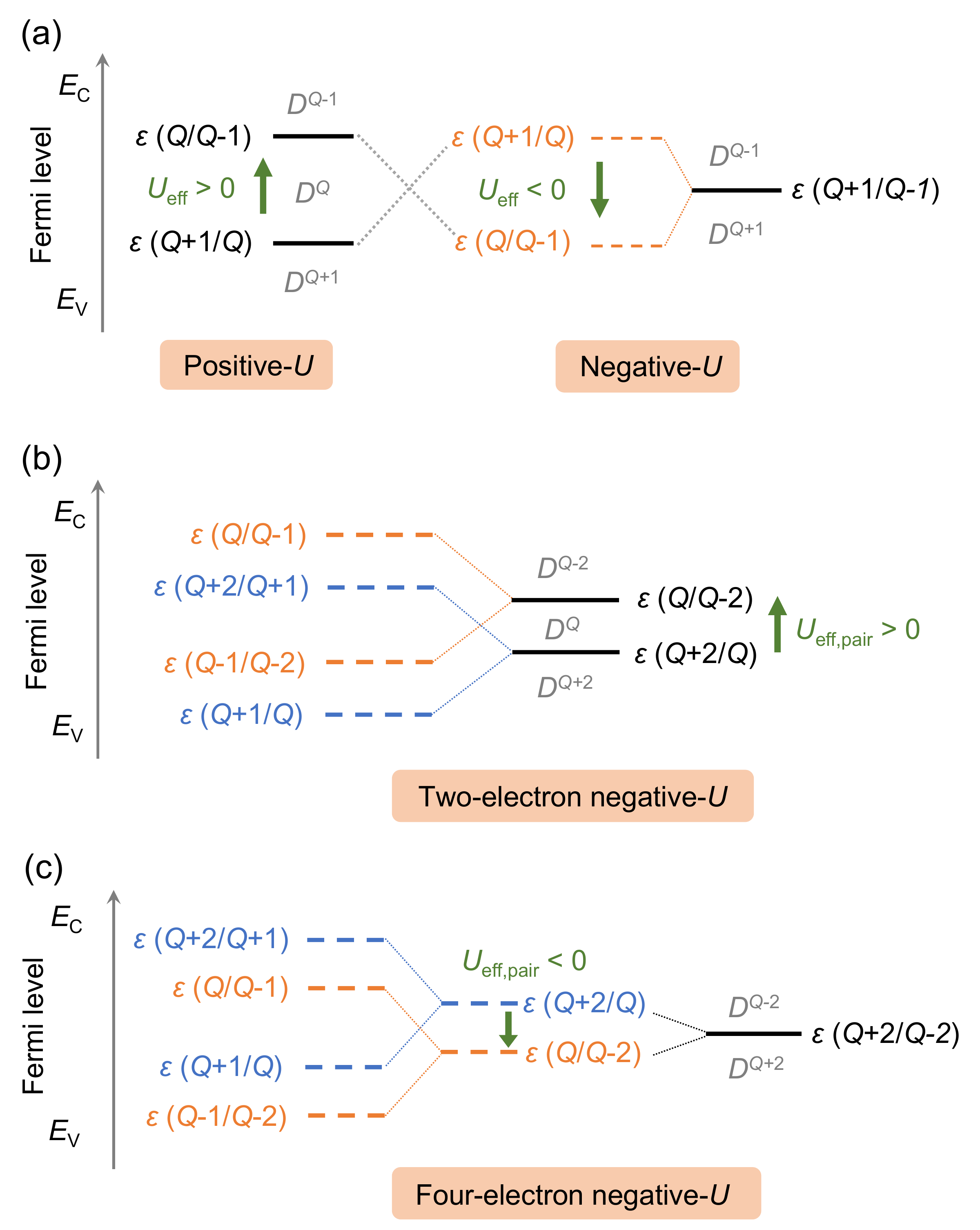}} \\
    \caption{Schematic diagrams of thermodynamic transition levels (TLs) for (a) positive- and negative-\textit{U} centers with three charge states, (b) two-electron negative-\textit{U} and (c) four-electron negative-\textit{U} vacancy centers in \ce{Sb2Se3} with five charge states. TLs are indicated by horizontal lines. \textit{D}(\textit{Q}) is the stable defect state under equilibrium conditions at a certain Fermi energy. \textit{E}$_\textrm{C}$, conduction band minimum. \textit{E}$_\textrm{V}$, valence band maximum.}
    \label{fig_tl}
\end{figure}

\textit{Many-electron negative-U behavior}.
The difference between positive- and negative-\textit{U} behaviors can be understood through the thermodynamic charge transition levels (TLs). The thermodynamic TL ($\varepsilon$(\textit{q}$_1$/\textit{q}$_2$)) is defined as the Fermi-level position where charge states \textit{q}$_1$ and \textit{q}$_2$ have the same defect formation energies:
\begin{equation}
\varepsilon(\textit{q}_1/\textit{q}_2)=\frac{\Delta E^f_{q_1}(\textit{E}_\textit{F}=0)-\Delta E^f_{q_2}(\textit{E}_\textit{F}=0)}{\textit{q}_2-\textit{q}_1}
\end{equation}
Let us first assume one defect which can be charged \textit{Q}+1, \textit{Q} or \textit{Q}-1 depending on the position of the Fermi level, where \textit{Q} is an arbitrary reference. 
Two TLs of $\varepsilon$(\textit{Q}+1/\textit{Q}) and $\varepsilon$(\textit{Q}/\textit{Q}-1) which define the borders between different charge states are thus considered. The energy separation of the $\varepsilon$(\textit{Q}+1/\textit{Q}) and $\varepsilon$(\textit{Q}/\textit{Q}-1) levels from the \ac{CBM} are the thermodynamic binding energies of the first and second electron to the defect, considering the capture process: \textit{D}$^{\textit{Q}+1}$ + 2\textit{e}$^-$ \textrightarrow { }\textit{D}$^{\textit{Q}}$ + \textit{e}$^-$ \textrightarrow { }\textit{D}$^{\textit{Q}-1}$. 
The energy difference between these two levels therefore determines the effective correlation energy \textit{U}$_\textrm{eff}$ (cost of trapping an additional charge carrier):
\begin{equation}
\textit{U}_\textrm{eff}=\varepsilon(\textit{Q}/\textit{Q}-1)-\varepsilon(\textit{Q}+1/\textit{Q})
\end{equation}
Consequently, the sign of \textit{U}$_\textrm{eff}$ depends on the ordering of these TLs. For a typical positive-\textit{U} center (\textit{U}$_\textrm{eff}$ \textgreater{ }0) (left-hand side of Fig. \ref{fig_tl}(a)), the position of $\varepsilon$(\textit{Q}+1/\textit{Q}) lies below $\varepsilon$(\textit{Q}/\textit{Q}-1), and so \textit{D}(\textit{Q}+1), \textit{D}(\textit{Q}) and \textit{D}(\textit{Q}-1) are each thermodynamically stable for some Fermi level. 
On the other hand, for a negative-\textit{U} center (\textit{U}$_\textrm{eff}$ \textless{ }0) (right-hand side of Fig. \ref{fig_tl}(a)), the ordering of $\varepsilon$(\textit{Q}+1/\textit{Q}) and $\varepsilon$(\textit{Q}/\textit{Q}-1) is inverted, resulting in a metastable \textit{D}(\textit{Q}) which would spontaneously emit charge and decay into \textit{D}(\textit{Q}+1) or \textit{D}(\textit{Q}-1), depending on \textit{E}$_F$. $\varepsilon$(\textit{Q}+1/\textit{Q}) and $\varepsilon$(\textit{Q}/\textit{Q}-1) are no longer thermodynamically stable TLs, and a single $\varepsilon$(\textit{Q}+1/\textit{Q}-1) level appears at the midpoint of these metastable TLs, such that:
\begin{equation}
\varepsilon(\textit{Q}+1/\textit{Q}-1)=\frac{\varepsilon(\textit{Q}+1/\textit{Q})+\varepsilon(\textit{Q}/\textit{Q}-1)}{2}
\end{equation}

Then let us move on to vacancy defects in \ce{Sb2Se3}. 
Since the thermodynamically stable TLs within the band gap are (+2/0) and (0/-2), or (+2/-2) for V$_\textrm{Se}$, and (+1/-3) for V$_\textrm{Sb}$ (Fig. \ref{fig_structure}(b)), five charge states (+1, 0, -1, -2 and -3 for V$_\textrm{Sb}$, and +2, +1, 0, -1 and -2 for V$_\textrm{Se}$) and four corresponding single-electron TLs are possible for both V$_\textrm{Sb}$ and V$_\textrm{Se}$. 
%The specific TL positions (given in Fig. S1, Tables S1 and S2) vary depending on the vacancy sites, but the relative ordering of the one-electron TLs is consistent between defects of the same type.
%-- except for 4-electron negative-U V$_\textrm{Se(2)}$ and 2-electron negative-U V$_\textrm{Se(1)}$ \& V$_\textrm{Se(3)}$.
%The specific values of TLs (given in Table S1 and S2) show a wide range depending on different vacancy types and different sites, but the trend between different TLs is similar for the same type (i.e. two- or four-electron) of negative-\textit{U} behavior. 
The single-electron TL positions are given in Fig. S1, Tables S1 and S2, with the relative ordering matching that of Fig. \ref{fig_tl}(b) for the two-electron negative-\textit{U} centers and Fig. \ref{fig_tl}(c) for four-electron centers. Here, \textit{Q}+2, \textit{Q}+1, \textit{Q}, \textit{Q}-1 and \textit{Q}-2 are used for convenience to explain the general pattern, where \textit{Q} equals -1 for V$_\textrm{Sb}$ and 0 for V$_\textrm{Se}$. 
As shown in Fig. \ref{fig_structure}(b) and depicted in Fig. \ref{fig_tl}(b) and (c), for all cases in V$_\textrm{Sb}$ and V$_\textrm{Se}$, the transitions of \textit{D}$^{\textit{Q}+2}$ \textrightarrow { }\textit{D}$^{\textit{Q}+1}$ \textrightarrow { }\textit{D}$^{\textit{Q}}$ (shown in blue dashed lines in Fig. \ref{fig_tl}) and \textit{D}$^{\textit{Q}}$ \textrightarrow { }\textit{D}$^{\textit{Q}-1}$ \textrightarrow { }\textit{D}$^{\textit{Q}-2}$ (shown in orange dashed lines) show negative-\textit{U} ordering and thus a two-electron transition for both \textit{D}$^{\textit{Q}+2}$ \textrightarrow { }\textit{D}$^{\textit{Q}}$ and \textit{D}$^{\textit{Q}}$ \textrightarrow { }\textit{D}$^{\textit{Q}-2}$.
Here for such cases with double two-electron TLs, we define an effective electron-\emph{pair} correlation energy \textit{U}$_\textrm{eff,pair}$ -- analogous to \textit{U}$_\textrm{eff}$ for single-electron correlation -- which corresponds to the energy cost of trapping an additional charge carrier pair:
\begin{equation}
\textit{U}_\textrm{eff,pair}=\varepsilon(\textit{Q}/\textit{Q}-2)-\varepsilon(\textit{Q}+2/\textit{Q})
\end{equation}
The relative positions of the $\varepsilon$(\textit{Q}/\textit{Q}-2) and $\varepsilon$(\textit{Q}+2/\textit{Q} negative-\textit{U} levels then dictate whether the defect is a two-electron or four-electron center. If $\varepsilon$(\textit{Q}+2/\textit{Q}) lies below $\varepsilon$(\textit{Q}/\textit{Q}-2) (Fig. \ref{fig_tl}(b)), a center with two typical two-electron negative-\textit{U} levels (\textit{U}$_\textrm{eff,pair}$ \textgreater { 0}) is formed. This is the case for V$_\textrm{Se(1)}$ and V$_\textrm{Se(3)}$ with \textit{U}$_\textrm{eff,pair}$ of 0.27 and 0.11 eV, respectively.
If on the other hand the ordering of these two levels are inverted once again (Fig. \ref{fig_tl}(c)), the defect would show a four-electron transfer behavior  (\textit{U}$_\textrm{eff,pair}$ \textless{ 0}) with the largest binding energy for the fourth electron, resulting in only two thermodynamically stable charge states \textit{D}$^{\textit{Q}+2}$ and \textit{D}$^{\textit{Q}-2}$ within the band gap.
This is the case for V$_\textrm{Sb(1)}$, V$_\textrm{Sb(2)}$ and V$_\textrm{Se(2)}$, with negative \textit{U}$_\textrm{eff,pair}$ values of -0.28, -0.09 and -0.04 eV respectively, thus showing four-electron negative-\textit{U} behavior.

Substantial structural relaxation is necessary to realize negative-\textit{U} behavior\cite{boer2018semiconductor,anderson1975model}. 
While a four-electron negative-\textit{U} transition is found for both sites in V$_\textrm{Sb}$, in V$_\textrm{Se}$ it depends on the Se site. Only V$_\textrm{Se(2)}$ shows a four-electron negative-\textit{U} behavior and the other sites show a two-electron one.
The two- or four-electron negative-\textit{U} behavior (i.e. the sign of \textit{U}$_\textrm{eff,pair}$) in V$_\textrm{Se}$ closely depends on the thermodynamic stability of the middle point \textit{D}$^{\textit{Q}}$ (i.e. the neutral defect). To quantify the effect of structural deformation on defect stability, we define the relaxation energy as the energy difference between unrelaxed and relaxed defect configurations. Relaxation energies of 0.33, 0.23 and 0.72 eV are obtained for V$^{0}_{\textrm{Se(1)}}$, V$^{0}_{\textrm{Se(2)}}$ and V$^{0}_{\textrm{Se(3)}}$ respectively, showing the opposite trend to the final formation energies; $E_f$(V$^{0}_{\textrm{Se(2)}}$) \textgreater{} $E_f$(V$^{0}_{\textrm{Se(1)}}$) \textgreater{} $E_f$(V$^{0}_{\textrm{Se(3)}}$) (as shown in Fig. S1). The stabilization from atomic relaxation is smallest for V$^{0}_{\textrm{Se(2)}}$, leading to the metastable V$^{0}_{\textrm{Se(2)}}$ and thus four-electron negative-\textit{U} level, showing the key role of structural reconstruction in determining the negative-\textit{U} properties of these defect species.

%【CONCLUSION】
In summary, structural configurations and electronic properties of vacancy defects in \ce{Sb2Se3} have been systematically modeled using a global search procedure. 
Thermodynamic TLs show that both types of vacancy are amphoteric, with strong self-charge-compensation and Fermi level pinning. In particular, both vacancies show either two- or four-electron negative-\textit{U} behaviors depending on the atomic site, indicating the vacancies in \ce{Sb2Se3} are likely to emit or capture charge carriers in pairs. 
This behavior is closely linked to structural reconstruction in V$^{+}_{\textrm{Sb}}$ and V$^{2-}_{\textrm{Se}}$, which are stabilized by the formation of Se/Sb trimers to pair electrons and compensate dangling bonds. Crucially, this behavior is not captured by standard defect modeling procedures, requiring global structure searching methods and consideration of non-typical charge states. The ability to reconstruct and thus stabilize these unusual defect charge states is attributed to the flexibility of the quasi-\ac{1D} crystal structure of \ce{Sb2Se3} and the valence alternation of Se and Sb species.
Our finding of multi-electron negative-\textit{U} behavior in \ce{Sb2Se3} suggests that the response of these materials to illumination may not follow standard electron-hole recombination kinetics with the possibility of multiple successive capture events from a single defect type.

%【Acknowledgement】
For computational resources we are grateful to the UK Materials and Molecular Modelling Hub and the UK's HEC Materials Chemistry Consortium, which are funded by EPSRC (EP/T022213/1 and EP/L000202). X.W. acknowledges Imperial College London for a President's PhD Scholarship. S.R.K. acknowledges the EPSRC Centre for Doctoral Training in the Advanced Characterisation of Materials (EP/S023259/1) for a PhD studentship. 

%\newpage
\bibliography{2-ref}

\end{document}

\begin{acronym}
\acro{VBM}{valence band maximum}
\acro{PESs}{potential energy surfaces}
\end{acronym}

\newpage

\section*{S1. The workflow of defect geometry optimization}
The workflow of generating and optimizing defects is discussed below. ~\\
~\\
I. \textbf{Defect generation}

The initial defect configurations were generated by both ~\\
i) local bond distortion of the nearest
neighbour atoms around the vacancy site  ~\\
ii) random perturbation for all atoms in the supercell.
~\\
~\\
Local bond distortions of both compression and stretching between 0\% and 60\% with 10\% as an interval were employed, following the \textsc{ShakeNBreak}\cite{mosquera2022shakenbreak} defaults, which have been tested across a range of materials and found to be sufficient for identifying ground-state configurations\cite{mosquera2022identifying}.
~\\
The displacement \textit{d} for all atoms is randomly chosen from a normal distribution of a specified standard deviation $\sigma$,
\begin{equation}
d \leftarrow \frac{1}{\sigma\sqrt{2\pi}}\textrm{exp}(-\frac{d^2}{2\sigma^2})
\end{equation}
$\sigma$ = 0.15, 0.20 and 0.25 Å were tested for each case.

~\\
II. \textbf{Structural relaxation}
~\\
As discussed above, for each defect, 39 initial trial atomic configurations are generated using \textsc{ShakeNBreak}\cite{mosquera2022shakenbreak}. In order to reduce the computational cost, we first performed hybrid optimisations with low accuracy (gamma \textit{k}-point) to identify the ground-state configuration for each defect, then the pre-converged structures with lowest energies were used as initial structures for further hybrid optimization with high accuracy (a denser \textit{k}-point mesh of 2 $\times$ 2 $\times$ 2) to get the final converged structures.

\newpage

\section*{S2. Transition level diagrams}
%~\\
\begin{figure*}[ht]
    \centering
    {\includegraphics[width=1.0\textwidth]{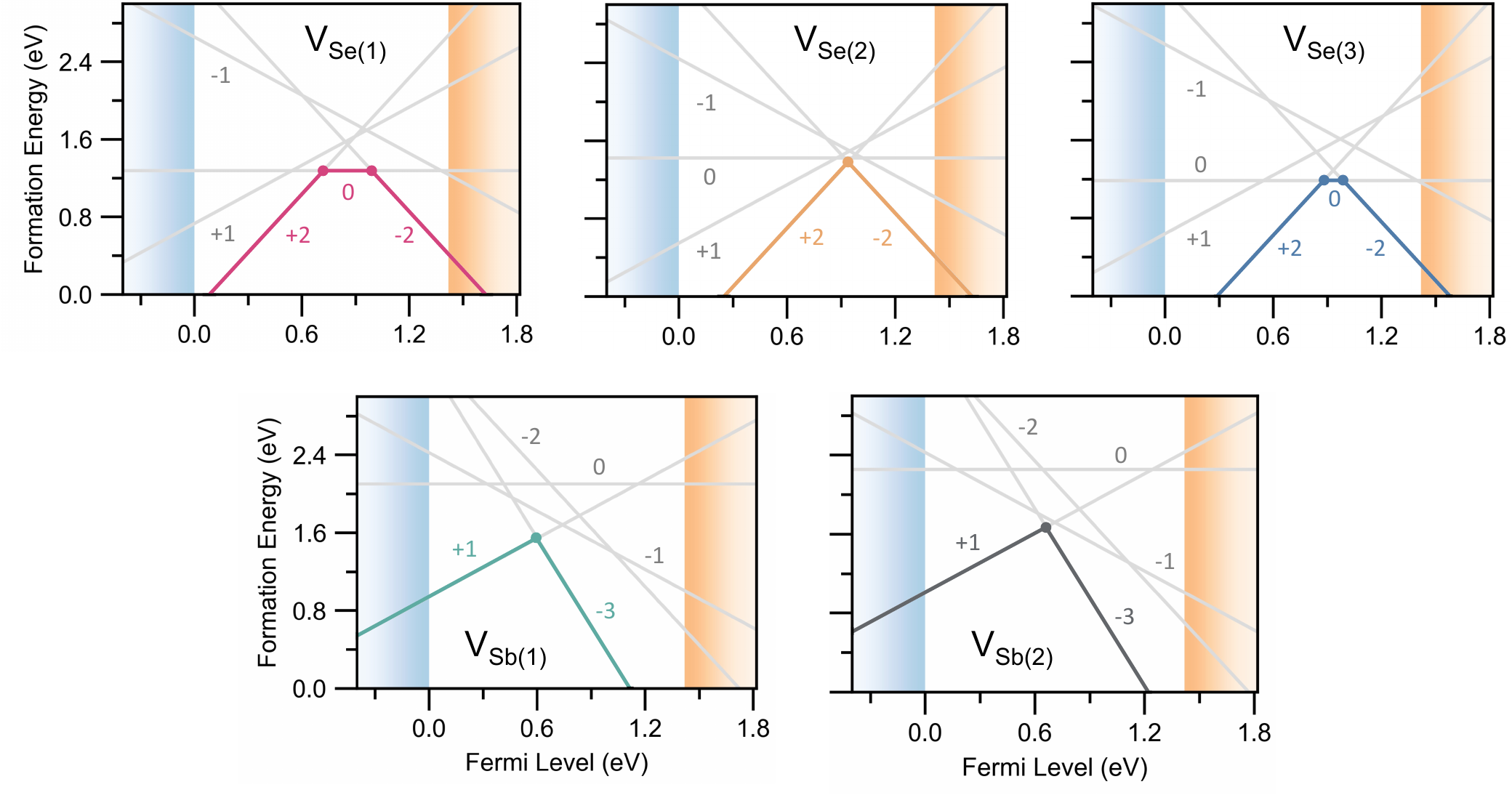}} \\
    \caption{Defect formation energy diagrams of V$_\textrm{Se}$ and V$_\textrm{Sb}$ including all thermodynamically metastable and stable states under Sb-rich equilibrium growth condition. Charge states are shown adjacent to lines. The thermodynamically metastable states are shown in grey, while the stable states are shown in bold. Valence band maximum (VBM) in blue, set to \SI{0}{eV}, and conduction band in orange.}
    \label{fig_all_tl}
\end{figure*}

\begin{table}[]
\centering
 \caption{Single-electron transition levels (eV) of V$_\textrm{Sb}$. The energy is referenced to \ac{VBM} which is set to 0. The metastable charge states are indicated with a superscript *. }
 \label{tab_tl_vsb}
\begin{tabular}{lclc}
 \hline
 \multicolumn{2}{c}{V$_\textrm{Sb(1)}$} & \multicolumn{2}{c}{V$_\textrm{Sb(2)}$} \\  \hline
\textit{\textepsilon} $(1/0^*)$ & 1.161 & \textit{\textepsilon} $(1/0^*)$ & 1.243 \\
\textit{\textepsilon} $(0^*/-1^*)$ & 0.320 & \textit{\textepsilon} $(0^*/-1^*)$ & 0.172 \\
\textit{\textepsilon} $(-1^*/-2^*)$ & 1.023 & \textit{\textepsilon} $(-1^*/-2^*)$ & 1.113 \\
\textit{\textepsilon} $(-2^*/-3)$ & -0.101 & \textit{\textepsilon} $(-2^*/-3)$ & 0.117 \\  \hline
\end{tabular}
\end{table}

\FloatBarrier
\begin{table}[t!]
\centering
 \caption{Single-electron transition levels (eV) of V$_\textrm{Se}$. The energy is referenced to \ac{VBM} which is set to 0. The metastable charge states are indicated with a superscript *.}
 \label{tab_tl_vse}
\begin{tabular}{lcclc}
 \hline
 & V$_\textrm{Se(1)}$ & V$_\textrm{Se(3)}$ & \multicolumn{2}{c}{V$_\textrm{Se(2)}$} \\  \hline
\textit{\textepsilon} $(2/1^*)$ & 0.896 & 1.214 & \textit{\textepsilon} $(2/1^*)$ & 1.047 \\
\textit{\textepsilon} $(1^*/0)$ & 0.544 & 0.545 & \textit{\textepsilon} $(1^*/0^*)$ & 0.875 \\
\textit{\textepsilon} $(0/-1^*)$ & 1.378 & 1.398 & \textit{\textepsilon} $(0^*/-1^*)$ & 1.019 \\
\textit{\textepsilon} $(-1^*/-2)$ & 0.603 & 0.573 & \textit{\textepsilon} $(-1^*/-2)$ & 0.817 \\ \hline
\end{tabular}
\end{table}
\FloatBarrier
~\\
\section*{S3. Sanity check for the cases with trimer formations}
However, it is worth noting that even though \textsc{ShakeNBreak} performs much better than the standard approach of simulating point defects, there is no guarantee that the final structure we get is necessarily the global minimum configuration, especially for systems with complex \ac{PESs} such as \ce{Sb2Se3}. 
~\\
~\\
After we obtained the relaxed defect configurations using \textsc{ShakeNBreak}, we found for some of the cases with Se/Sb trimer formations (i.e. V$_\textrm{Sb}^+$/V$_\textrm{Se}^{2-}$), the trimer was formed via the movement of neighboring Se/Sb atom towards the Sb/Se vacancy site. This phenomenon is more obvious in V$_\textrm{Sb(1)}^+$ and V$_\textrm{Sb(2)}^+$, but also happens in V$_\textrm{Se(3)}^{2-}$.
Considering that other inequivalent Se/Sb atoms could also possibly move to the vacancy site, as a sanity check, we further tested different initial configurations for those cases and compared their energies after structural relaxation. 
For each case, the initial configuration was generated by intentionally moving the possible inequivalent neighboring Se/Sb atom to the vacancy site, and then structural optimisation with gamma \textit{k}-point was performed. 
The results are shown in Table \ref{tab_c_vsb}. The distance between Se(2) and Sb(2) is reasonably large (4.71 Å), so this case was not considered.
~\\
\FloatBarrier
\begin{table}[t!]
\centering
\caption{The energies (eV) of different initial defect configurations in V$_\textrm{Sb}^+$ and V$_\textrm{Se}^{2-}$. The energy is referenced to the most stable case which is set to 0. Se(x) or Sb(x) refers to the neighboring Se/Sb atom which moves to the Sb/Se vacancy site after relaxation.}
 \label{tab_c_vsb}
\begin{tabular}{ccccc}
\hline
 & V$_\textrm{Sb(1)}^+$ & V$_\textrm{Sb(2)}^+$ & & V$_\textrm{Se(3)}^{2-}$ \\  \hline
$\textrm{Se(1)}$ & 0.62 & 0.00 &$\textrm{Sb(1)}$& 0.51 \\
$\textrm{Se(2)}$ & 0.00 & - &$\textrm{Sb(2)}$& 0.00 \\ 
$\textrm{Se(3)}$ & 0.13 & 0.18 & & \\   \hline
\end{tabular}
\end{table}
\FloatBarrier

~\\
%\clearpage
%\newpage
%\section*{References}
\bibliography{2-SI}